\documentclass[prl,twocolumn,twoside,superscriptaddress,nofootinbib]{revtex4}

\usepackage{graphicx} 
\usepackage[pdfauthor={Matthew Neeley}, pdftitle={Generation of Three-Qubit Entangled States using Superconducting Phase Qubits}]{hyperref} 
\usepackage{multirow} 
\usepackage{dcolumn} 

\newcommand{\ket}[1]{\ensuremath{\left|#1\right\rangle}}
\newcommand{\bra}[1]{\ensuremath{\left\langle#1\right|}}

\newcommand{\GHZ}{\ket{\mathrm{GHZ}}}
\newcommand{\braGHZ}{\bra{\mathrm{GHZ}}}
\newcommand{\W}{\ket{\mathrm{W}}}
\newcommand{\braW}{\bra{\mathrm{W}}}
\newcommand{\iSWAP}{iSWAP}

\renewcommand{\Im}{\ensuremath{\mathrm{Im}}}

\newcommand{\UCSB}{Department of Physics, University of California, Santa Barbara, CA 93106, USA}
\newcommand{\NEC}{Green Innovation Research Laboratories, NEC Corporation, Tsukuba, Ibaraki 305-8501, Japan}

\begin{document}

\title{Generation of Three-Qubit Entangled States using Superconducting Phase Qubits}

\author{M. Neeley}\affiliation{\UCSB}
\author{R. C. Bialczak}\affiliation{\UCSB}
\author{M. Lenander}\affiliation{\UCSB}
\author{E. Lucero}\affiliation{\UCSB}
\author{M. Mariantoni}\affiliation{\UCSB}
\author{A. D. O'Connell}\affiliation{\UCSB}
\author{D. Sank}\affiliation{\UCSB}
\author{H. Wang}\affiliation{\UCSB}
\author{M. Weides}\affiliation{\UCSB}
\author{J. Wenner}\affiliation{\UCSB}
\author{Y. Yin}\affiliation{\UCSB}
\author{T. Yamamoto}\affiliation{\UCSB}\affiliation{\NEC}
\author{A. N. Cleland}\affiliation{\UCSB}
\author{J. M. Martinis}\affiliation{\UCSB}

\begin{abstract}
Entanglement is one of the key resources required for quantum computation\cite{Nielsen2000}, so experimentally creating and measuring entangled states is of crucial importance in the various physical implementations of a quantum computer\cite{Ladd2010}.  In superconducting qubits\cite{Clarke2008}, two-qubit entangled states have been demonstrated and used to show violations of Bell's Inequality\cite{Ansmann2009} and to implement simple quantum algorithms\cite{DiCarlo2009}.  Unlike the two-qubit case, however, where all maximally-entangled two-qubit states are equivalent up to local changes of basis, three qubits can be entangled in two fundamentally different ways\cite{Duer2000}, typified by the states $\GHZ = (\ket{000} + \ket{111})/\sqrt{2}$ and $\W = (\ket{001} + \ket{010} + \ket{100})/\sqrt{3}$.  Here we demonstrate the operation of three coupled superconducting phase qubits\cite{McDermott2005} and use them to create and measure $\GHZ$ and $\W$ states\footnote{Independently, entanglement created between three superconducting transmon qubits is being reported in a simultaneous publication\cite{DiCarlo2010}.}.  The states are fully characterized using quantum state tomography\cite{Steffen2006} and are shown to satisfy entanglement witnesses\cite{Acin2001}, confirming that they are indeed examples of three-qubit entanglement and are not separable into mixtures of two-qubit entanglement.
\end{abstract}

\maketitle

\addtocounter{footnote}{1} 

In order to create arbitrary entangled states or perform arbitrary computations, a quantum computer must implement a set of universal gates\cite{Nielsen2000}, typically taken to be a two-qubit gate such as controlled-NOT (CNOT) plus single qubit rotations\cite{Barenco1995}.  Alternately, universality is possible using a three-qubit gate such as the Toffoli gate\cite{Shi2003, Lanyon2009, Monz2009}.  Three-qubit gates are also important in such applications as quantum error-correction\cite{Cory1998} and they can simplify some quantum circuits\cite{Lanyon2009}.  Because superconducting phase qubits can be coupled simply by connecting them with a capacitor\cite{McDermott2005}, we can design multi-qubit interactions that directly generate multi-qubit gates\cite{Galiautdinov2008}, rather than building them up from more elementary two-qubit gates.  For creating the two types of three-qubit entanglement we employ both approaches, using two-qubit gates for the $\GHZ$ protocol, but using a more efficient entangling protocol for $\W$ based on a single three-qubit gate.

\begin{figure}
\includegraphics[scale=0.95, trim=0in 1.85in 0in 0in, clip=true]{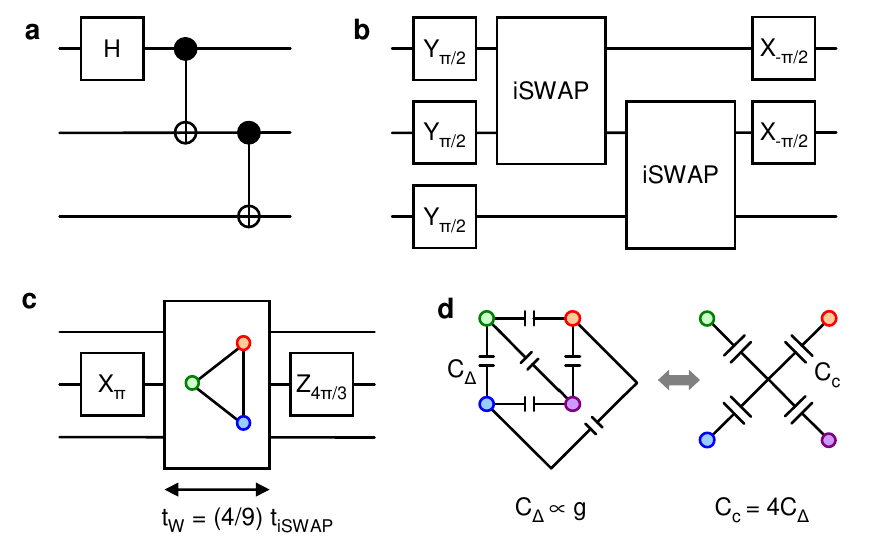}
\caption{\label{protocols} Protocols for generating entangled states.  {\bf a}, Quantum circuit for generating $\GHZ$ using CNOT gates.  {\bf b}, Quantum circuit for $\GHZ$ that has been ``recompiled'' to use \iSWAP\ gates, which are directly generated by capacitive coupling in the phase qubit.  These two circuits are not fully equivalent, but they both produce $\GHZ$ when operating on the ground state as input.  {\bf c}, Circuit to generate $\W$ using a single entangling step with simultaneous coupling between all three qubits.  The entangling operation is turned on for a time $t_{\mathrm{W}} = (4/9) t_{\mathrm{\iSWAP}}$ where $t_{\mathrm{\iSWAP}}$ is the time needed to complete an \iSWAP\ gate between two qubits.  {\bf d}, Capacitive coupling network to achieve symmetric coupling between all pairs of qubits (left), and simplified equivalent circuit using coupling to a central island (right).  The complete network on the left requires six capacitors, and the coupling strength $g$ is proportional to the qubit-qubit capacitance $C_{\Delta}$.  In the equivalent circuit on the right, the same coupling strength is attained by scaling the capacitors to $C_c = 4 C_{\Delta}$, but now only four capacitors are required and the circuit can be easily laid out symmetrically on a chip.}
\end{figure}

The $\GHZ$ protocol\cite{Wei2006, Matsuo2007} is shown as a quantum circuit diagram in Figure \ref{protocols}a.  Starting in the ground state $\ket{000}$, a rotation is applied to qubit $A$ to create the superposition $(\ket{000} + \ket{100})/\sqrt{2}$.  Next, a CNOT gate is applied to flip qubit $B$ conditioned on qubit $A$, resulting in the state $(\ket{000} + \ket{110})/\sqrt{2}$.  Finally a second CNOT is applied to flip qubit $C$ conditioned on $B$, resulting in the desired state $\GHZ$.  As is typical with quantum circuits, this is written in terms of CNOT gates which take a simple form in the qubit basis.  In our system however, a more natural universal gate is the so-called \iSWAP\ gate\cite{Geller2010} which is generated directly by applying the available coupling interaction $H_{\mathrm{int}}^{AB} = (\hbar g/2)(\sigma_x^A\sigma_x^B + \sigma_y^A\sigma_y^B)$ for time $t_{\mathrm{\iSWAP}} = \pi/2g$, where $g$ is the coupling strength.  The $\GHZ$ protocol can be ``recompiled'' in terms of this gate to obtain the circuit shown in Figure \ref{protocols}b.

The protocol to generate a $\W$-state, shown in Figure \ref{protocols}c, is based on two features of the $\W$-state: it is symmetric with respect to permutations of the qubits, and it is a superposition of three states each with one qubit excited.  Thus, generating the state requires ``sharing'' a single excitation symmetrically among three qubits.  This is done by first applying a $\pi$-pulse to qubit $B$ to excite it with one photon and create the state $\ket{010}$.  Then the qubits are entangled by turning on an equal interaction between all pairs $H_{\mathrm{int}} = H_{\mathrm{int}}^{AB} + H_{\mathrm{int}}^{AC} + H_{\mathrm{int}}^{BC}$ for time $t_{\mathrm{W}} = (4/9) t_{\mathrm{\iSWAP}}$.  The interaction causes the excitation to be distributed among the qubits, and at time $t_{\mathrm{W}}$ the system is left in an equal superposition state, as desired.  A final $Z$-rotation can then be applied to correct the phase of qubit $B$, though this does not affect the entanglement of the state.  This protocol requires only a single entangling operation, and the interaction is only applied for a short time, shorter even than the characteristic time for two-qubit gates in the system.  This yields a highly efficient state-generation protocol based on the multi-qubit gate generated by $H_{\mathrm{int}}$.

To allow for future expansion beyond the present work, the sample was designed with four qubits, so that the coupling network for the desired symmetric coupling between all pairs of qubits is as shown in Figure \ref{protocols}d (left).  The design can be simplified by transforming the coupling network into an equivalent circuit (right) with each qubit coupled capacitively to a central ``island''.  This simplified design is easier to lay out symmetrically on chip and requires only $N$ capacitors to couple $N$ qubits, rather than the $N(N-1)/2$ capacitors in the complete network.

\begin{figure}
\includegraphics[scale=0.95]{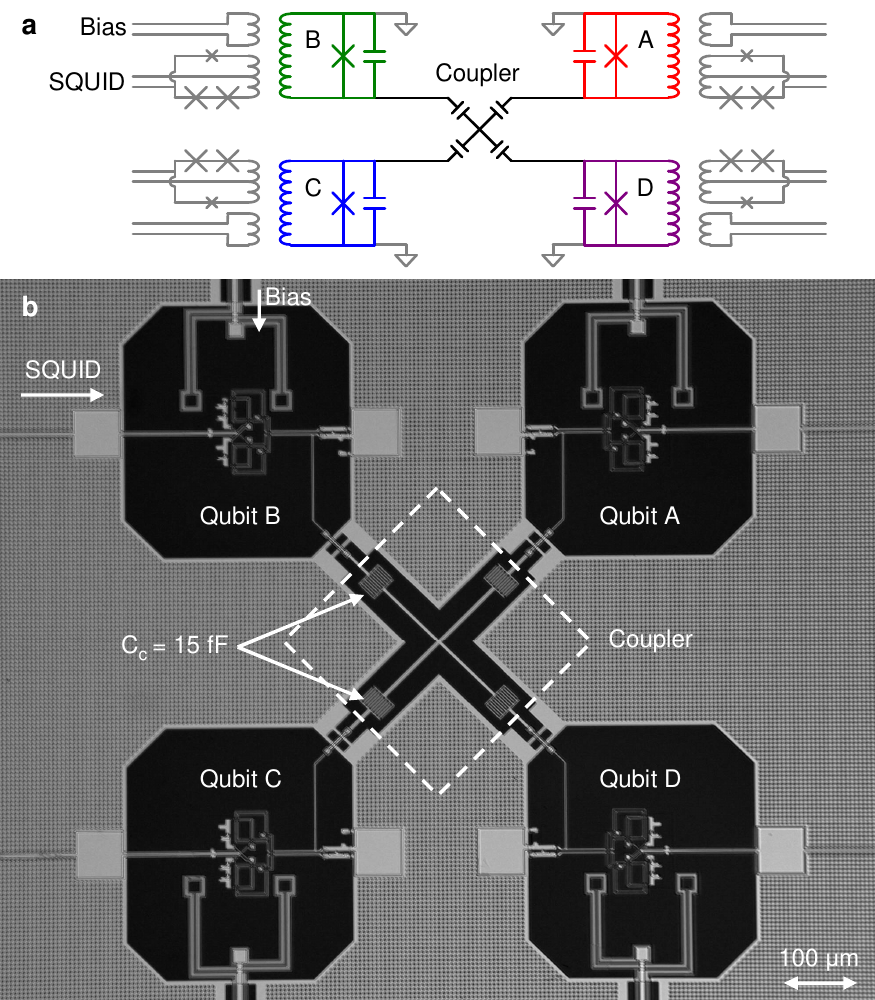}
\caption{\label{sample} Device description and operation. {\bf a}, Schematic of coupled-qubit circuit.  Each qubit is controlled individually by a flux bias line which sets the DC operating point, and provides quasi-DC pulses for tuning the qubits in and out of resonance and AC (microwave) control signals for qubit rotations.  In addition, each qubit is coupled to a superconducting quantum interference device (SQUID) for readout of the qubit state.  The qubits are capacitively coupled to the central island, which results in symmetric coupling between all pairs of qubits.  {\bf b}, Photomicrograph of the sample, fabricated with aluminum (light areas) on sapphire substrate (dark areas).  The coupler is the cross-shaped structure in the center, and the simplicity of this design is evident in the straightforward correspondence between the schematic and the completed device.  The entire sample is mounted in a superconducting aluminum box and cooled to $25\,\mathrm{mK}$ in a dilution refrigerator.}
\end{figure}

Figure \ref{sample}a shows the complete schematic of the device with four phase qubits connected by the capacitive island coupler.  Each qubit is individually controlled by a bias coil which sets the operating flux bias and also carries microwave pulses for manipulating and measuring the qubit state.  In addition, each qubit is coupled to an on-chip superconducting quantum interference device (SQUID) for state readout.  Figure \ref{sample}b shows a micrograph of the fabricated device, made from aluminum films on sapphire substrate with Al/AlO$_x$/Al Josephson junctions.  The completed device is mounted in a superconducting aluminum sample holder and cooled in a dilution refrigerator to $\sim 25\,\mathrm{mK}$.  Bringup and calibration of the multiqubit device are similar to previous works\cite{Lucero2008, Neeley2009}.  Although the coupling capacitors are fixed, the effective interaction can be controlled by tuning the qubits into resonance at $f_B = 6.55\,\mathrm{GHz}$ (coupling ``on'') or by detuning $A$ and $C$ to $\pm 250\,\mathrm{MHz}$ (coupling ``off'')\cite{Hofheinz2009}.  The measured coupling strengths were found to be within $5\%$ of $12.5\,\mathrm{MHz}$ for each pair of qubits.  Importantly, all qubits can be brought into resonance simultaneously, as required for the $\W$ protocol, or two qubits can be tuned into resonance with the third detuned, as required for the \iSWAP\ gates in the $\GHZ$ protocol.

Capacitive coupling as used here is simple and well-understood but is subject to measurement crosstalk\cite{McDermott2005, Ansmann2009}, which can cause, for example, a state $\ket{001}$ to be erroneously read out as $\ket{011}$, $\ket{101}$ or even $\ket{111}$.  This crosstalk affects measured probabilities of all excited-state populations, however it has no effect on the ``null-result'' probability of measuring $\ket{000}$, since crosstalk can only act if at least one qubit is excited.  By measuring various subsets of qubits and recording the null-result probability for each subset, we are able to reconstruct the combined state occupation probabilities without any effect from measurement crosstalk\footnote{See supplementary information.}.

Figure \ref{timedomain} shows the time-evolution of the state occupation probabilities during the entangling protocols, using crosstalk-free measurement.  In the $\W$ protocol (Figure \ref{timedomain}a), one qubit is excited and then the symmetric interaction between all pairs of qubits is used to distribute that excitation among all three, as described above.  When the interaction time is chosen properly, the system reaches an equal superposition, and subsequently stays there while the interaction is off (Figure \ref{timedomain}b).  Because this one excitation is swapped among the various qubits, the state evolution during this protocol is clearly visible in the occupation probabilities as they evolve in time.

Figure \ref{timedomain}c shows the state occupation probabilities during the $\GHZ$ protocol, plotted in segments corresponding to the stages of the protocol as indicated.  The initial rotations create an equal superposition of all qubit states, with all probabilities converging on $1/8$.  The effect of the two \iSWAP\ gates is then primarily to adjust the phases of the various components of the superpositions, so that in the final rotation constructive interference causes $\ket{000}$ and $\ket{111}$ to be populated, while all other states are depopulated.  The occupation probabilities behave as expected, but most of the state evolution is hidden in the phase information not captured by these probability measurements.

\begin{figure}
\includegraphics[scale=0.95, trim=0in 0.9in 0in 0in, clip=true]{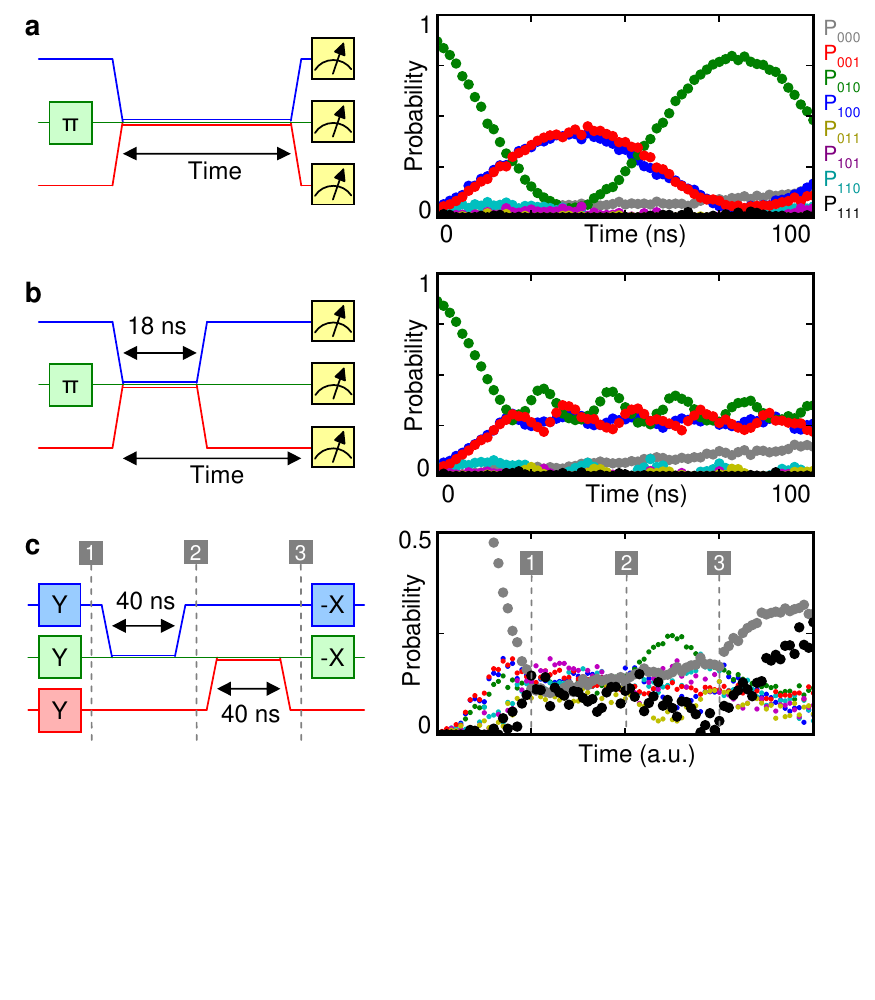}
\caption{\label{timedomain} Generation of entangled states in the time domain.  In each row, the left panel shows the pulse sequence with time on the horizontal axis and qubit frequency on the vertical axis.  The right panel shows the measured state occupation probabilities as a function of time during each sequence.  {\bf a}, To characterize the three-qubit interaction, all qubits are initially detuned and qubit $B$ is excited with a $\pi$-pulse.  The qubits are then tuned into resonance to turn on the interaction for some time, then detuned and measured.  During the interaction, the excitation from qubit $B$ ($\ket{010}$) is swapped to qubits $A$ and $C$ ($\ket{100}$ and $\ket{001}$), then back again.  Probabilities $P_{100}$ and $P_{001}$ are nearly equal throughout the entire sequence, indicating that the coupling is nearly symmetric, as desired.  At the first crossing point where the three probabilities are equal, the system is in a $\W$-state.  {\bf b}, The coupling is turned on until the crossing point is reached and then the qubits are detuned, leaving the system in a $\W$-like state, up to phase rotations due to the detunings.  The small residual oscillations visible after the qubits have been detuned are due to interactions with $\ket{2}$ and higher energy levels of the phase qubits.  {\bf c}, The $\GHZ$ sequence is a direct translation of the circuit shown in Figure \ref{protocols}b, with the \iSWAP\ gates implemented by tuning the qubits pairwise into resonance for time $t_{\mathrm{\iSWAP}} = 40\,\mathrm{ns}$.  The occupation probabilities on the right are plotted versus time in each marked stage of the sequence.  After creating the initial superposition (1), the two \iSWAP\ gates change the phases of the various components of the state, with little effect on the populations (1-2, 2-3).  During the final rotation, constructive interference populates $\ket{000}$ and $\ket{111}$, while destructive interference depopulates the other states.  For an ideal $\GHZ$ state, the probabilities $P_{000}$ and $P_{111}$ should approach $50\%$, though in the experiment this level is reduced due to the effects of decoherence and errors discussed in the text.}
\end{figure}

\begin{figure*}
\includegraphics[trim=0in 4.2in 0in 0in, clip=true]{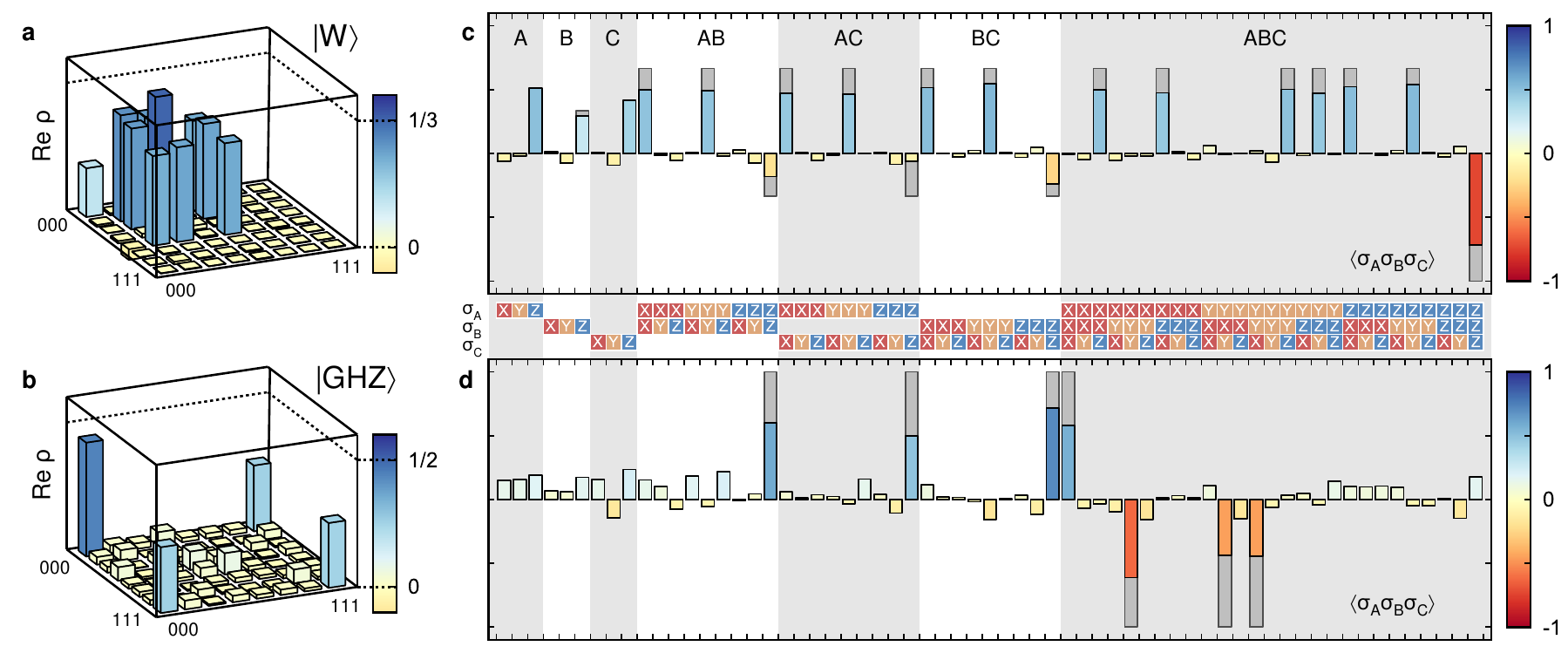}
\caption{\label{tomography} Quantum state tomography of $\GHZ$ and $\W$.  At left, the real parts of the measured density matrices $\rho_{\mathrm{W}}$ ({\bf a}) and $\rho_{\mathrm{GHZ}}$ ({\bf b}) are shown in the bar plots.  For both states the theoretical density matrix has vanishing imaginary part, and the measured imaginary parts (not shown) are also found to be small, with $|\Im\,\rho_{\mathrm{W}}| < 0.03$ and $|\Im\,\rho_{\mathrm{GHZ}}| < 0.10$ respectively.  At right, the Pauli set or generalized Stokes parameters are plotted for $\rho_{\mathrm{W}}$ ({\bf c}) and $\rho_{\mathrm{GHZ}}$ ({\bf d}).  The bars show expectation values of combinations of Pauli operators on one, two and three qubits, with theory in gray and experiment overlayed in color.  The same state information is contained in both representations, but the Pauli sets clearly show the differences between $\W$-type and $\GHZ$-type entanglement.  In addition to the three-qubit correlation terms, the $\W$-state has two-qubit correlations because tracing out one qubit from a $\W$-state still leaves the others partially entangled.  The fidelity is $F_{\mathrm{W}} = 0.78$.  For $\GHZ$, the two-qubit correlations other than the trivial $ZZ$-type are absent because tracing out one qubit leaves the others in a completely mixed state.  The fidelity is $F_{\mathrm{GHZ}} = 0.62$ and the state is also found to violate the Mermin-Bell inequality\cite{Mermin1990b, Pan2000}.}
\end{figure*}

To fully characterize the quantum states created by the entangling protocols, including the phase information, we perform Quantum State Tomography (QST) by applying various combinations of single-qubit rotations before measurement.  The density matrix is extracted from the measured data using maximum likelihood estimation (MLE) to find the state that best fits the data while also satisfying the physicality constraints that it be Hermitian positive semi-definite with unit trace.  Using this procedure, we extract $\rho_{\mathrm{W}}$ and $\rho_{\mathrm{GHZ}}$, shown respectively in Figures \ref{tomography}a and \ref{tomography}b.  Comparing the measured states with theory, we find fidelities $F_{\mathrm{W}} = \braW\rho_{\mathrm{W}}\W = 0.78$ and $F_{\mathrm{GHZ}} \equiv \braGHZ\rho_{\mathrm{GHZ}}\GHZ = 0.62$.

To understand the significance of the measured fidelities, we compare these results to entanglement witness operators that detect three-qubit entanglement.  Three-qubit entanglement is witnessed\cite{Acin2001} for the $\W$-state provided that $F_{\mathrm{W}} > 2/3$, and for the $\GHZ$-state provided that $F_{\mathrm{GHZ}} > 1/2$.  Both inequalities are satisfied by the respective measured density matrices, indicating that they are genuine three-qubit entangled states that cannot be decomposed into mixtures of separable states.  In addition, $\rho_{\mathrm{GHZ}}$ is found to violate the Mermin-Bell inequality\cite{Mermin1990b}, as predicted by quantum mechanics but disallowed by the classical assumptions of local reality\footnote{See supplementary information.}.  The violation is not loophole-free due to use of the crosstalk-free measurement protocol rather than a simultaneous measurement protocol\cite{Ansmann2009}, but it is nonetheless an indicator of genuine three-qubit entanglement.

The lower fidelity of $\GHZ$ compared to $\W$ is due to two main factors: first, the $\GHZ$ sequence is longer because of the two \iSWAP\ gates; the sequence length is a substantial fraction of the dephasing time $T_2$ of the qubits, which is particularly harmful because the sequence relies on precise phase adjustment and interference to populate $\ket{000}$ and $\ket{111}$ while depopulating all other states.  Longer coherence times would improve this, as would stronger coupling to reduce the gate time.  Second, the presence of $\ket{2}$ and higher levels and the relatively small nonlinearity of the phase qubit cause errors due to transitions into higher excited states, for example $\ket{110}\rightarrow\ket{200}$.  These transitions can be ignored in the $\W$ protocol since they are inaccessible with only one excitation in the system, but they cause errors in the $\GHZ$ protocol since all qubit states are populated, including those with multiple excitations.  The effect of higher levels becomes particularly complicated in this experiment when using fixed capacitive coupling with detuning to turn off the interaction, due to spectral crowding from the higher qubit levels.  This highlights the need to replace frequency detuning with tunable coupling schemes, which are currently an active area of research.

In conclusion, we used superconducting phase qubits to generate both types of three-qubit entangled states, namely $\GHZ$ and $\W$.  In both cases, the created states violate entanglement witnesses that rule out biseparability, showing that these are genuine three-qubit entangled states.  This ability to couple three qubits and create entangled states with qualitatively different types of entanglement represents an important step toward scalable quantum information processing with superconducting devices.

{\small
Devices were made at the UCSB Nanofabrication Facility, a part of the NSF-funded National Nanotechnology Infrastructure Network. This work was supported by IARPA under grant W911NF-04-1-0204.  M.M. acknowledges support from an Elings Fellowship.

Correspondence and requests for materials should be addressed to J.M.M.~(email: martinis@physics.ucsb.edu).
}

\bibliography{entanglement}

\end{document}


\title{Generation of Three-Qubit Entangled States using Superconducting Phase Qubits: Supplementary Information}

\newcommand{\UCSB}{Department of Physics, University of California, Santa Barbara, CA 93106, USA}
\newcommand{\NEC}{Green Innovation Research Laboratories, NEC Corporation, Tsukuba, Ibaraki 305-8501, Japan}

\author{Matthew Neeley}\affiliation{\UCSB}
\author{Radoslaw C. Bialczak}\affiliation{\UCSB}
\author{M. Lenander}\affiliation{\UCSB}
\author{E. Lucero}\affiliation{\UCSB}
\author{M. Mariantoni}\affiliation{\UCSB}
\author{A. D. O'Connell}\affiliation{\UCSB}
\author{D. Sank}\affiliation{\UCSB}
\author{H. Wang}\affiliation{\UCSB}
\author{M. Weides}\affiliation{\UCSB}
\author{J. Wenner}\affiliation{\UCSB}
\author{Y. Yin}\affiliation{\UCSB}
\author{T. Yamamoto}\affiliation{\UCSB}\affiliation{\NEC}
\author{A. N. Cleland}\affiliation{\UCSB}
\author{John M. Martinis}\affiliation{\UCSB}

\maketitle 

\section{Materials and Methods}
The phase qubits in this device were designed to have critical current $I_0\approx 2$~$\mu$A, capacitance $C\approx 1$~pF, and inductance $L\approx 720$~pH.  The coupling capacitance was $C_c\approx 15$~pF, chosen to give coupling strength $2g/2\pi\approx 15$~MHz at a qubit frequency of $6$~GHz.  The fabrication process is the same as has been used in previous experiments\cite{Neeley2009}, using a sapphire substrate with superconducting Al films, AlO$_x$ tunnel junction, and a-Si:H dielectric for the qubit and SQUID shunt capacitors and for wiring crossovers.  Each qubit is controlled with a single line which provides DC flux bias, quasi-DC detuning and measurement pulses, and microwave state-rotation pulses.  Each qubit is coupled to a three-junction measurement SQUID for readout.

The microwave control signals are produced by a custom microwave arbitrary waveform generator (AWG), which has been described in the supplementary information to previous work\cite{Hofheinz2009}.  For single-qubit rotations we use $6$~ns full-width at half-max (FHWM) gaussian pulses.  To reduce errors in these operations due to the presence of the $\ket{2}$-state and the small nonlinearity of the phase qubit, the pulses include a quadrature modulation correction inspired by the scheme known as Derivative Removal by Adiabatic Gate (DRAG)\cite{Motzoi2009}.

\section{Bringup and Calibration}
To operate the sample, we begin by characterizing the operation of the individual qubits.  From basic single-qubit experiments, we find relaxation times and spin-echo dephasing times for the qubits, as shown in Table \ref{singlequbit}.  The single-qubit characterizations were performed with the qubits detuned by $\pm 250\,\mathrm{MHz}$ as indicated in the table, so that the coupling interaction is off since the detuning $\Delta$ is much larger than the coupling strength $\Delta/g \approx 20$.

\begin{table}
\begin{tabular}{c | c | c | c}
  qubit & $T_1$ (ns) & $T_{\mathrm{echo}}$ (ns) & $f_{10}$ (GHz) \\
\hline
  A & 460 & 270 & 6.2995 \\
  B & 460 & 300 & 6.5506 \\
  C & 450 & 390 & 6.7988 \\
\end{tabular}
\caption{\label{singlequbit} Single-qubit parameters.}
\end{table}

Turning the coupling interaction on requires tuning the qubits into resonance with each other for some time.  The required detuning pulses are calibrated pairwise between the qubits; for example, to calibrate the detuning pulse to couple $A$ with $B$, we first excite qubit $B$ with a $\pi$-pulse and then adjust the amplitude and length of the detuning pulse on $A$ to maximize the transfer of this excitation to $A$.  This pulse creates an iSWAP gate, as needed for the $\ket{\mathrm{GHZ}}$ protocol.  The process is repeated for each pair of qubits, giving swap times and coupling strengths as shown in Table \ref{coupling}.  The coupling strengths are within 5\% of each other and also quite close to the design value of $15\,\mathrm{MHz}$.  For the $\ket{\mathrm{W}}$-state protocol the interaction time is $t_{\mathrm{W}} = (4/9) t_{\mathrm{iSWAP}} \approx 18\,\mathrm{ns}$.

\begin{table}
\begin{tabular}{c | c | c}
  qubits & $t_{\mathrm{iSWAP}}$ (ns) & $2g/2\pi$ (MHz) \\
\hline
  AB & 40.3 & 12.4 \\
  AC & 40.9 & 12.2 \\
  BC & 38.8 & 12.9 \\
\end{tabular}
\caption{\label{coupling} Qubit-qubit coupling parameters.}
\end{table}

\section{Crosstalk-free measurement protocol}
The capacitive coupling used in this sample is subject to measurement crosstalk, in which the measurement of one qubit as being in state $\ket{1}$ can radiate energy into the circuit and cause other qubits to switch.  With only two coupled qubits, crosstalk can cause errors $\ket{01}\rightarrow\ket{11}$ and $\ket{10}\rightarrow\ket{11}$.  In previous work with coupled qubits\cite{Bialczak2010}, we have measured the probabilities of these two errors and then simply corrected the qubit measurement results to account for this crosstalk.  However with three qubits all coupled together, the effect of crosstalk is much more complicated because there are more possible errors, for example $\ket{001}\rightarrow\ket{011}$, $\ket{001}\rightarrow\ket{101}$, $\ket{001}\rightarrow\ket{111}$ and so on.  While all these error probabilities could in principle be measured and corrected for, we instead choose to measure in a way that is entirely insensitive to measurement crosstalk, thus eliminating the need for any correction or even any characterization of the crosstalk errors.

The essential element of this scheme is the fact that crosstalk does not affect the state $\ket{000}$ since no qubits are excited.  Hence when all the qubits are measured to find the probabilities $P_{000}, P_{001}, ... P_{111}$ of various outcomes, only the ``null-result'' probability $P_{000}$ of no qubits excited requires no crosstalk correction.  Now, consider the case when qubits $A$ and $B$ are measured but qubit $C$ is not.  We obtain four probabilities $P_{00x}, P_{01x}, P_{10x}, P_{11x}$, where the subscript $x$ indicates that in each case we have no information about the unmeasured qubit $C$.  As before, only the null-result probability $P_{00x}$ is unaffected by crosstalk, but note that $P_{00x} = P_{000} + P_{001}$ because there are two possibilities for the state of the unmeasured qubit $C$.  The two probabilities $P_{00x}$ and $P_{000}$ can be measured without crosstalk as just described; from these the third probability $P_{001}$ can be determined.

Continuing in this manner, we can reconstruct the complete set of occupation probabilities without crosstalk by repeating the experiment $2^3 - 1 = 7$ times, each time measuring only a certain subset of the qubits and recording only the null-result probability for that subset (for the degenerate case in which no qubits are measured, we have $P_{xxx}=1$).  This gives the following set of measured null-results $\mathbf{P}_{\mathrm{null}} = (P_{000}, P_{00x}, P_{0x0}, P_{0xx}, P_{x00}, P_{x0x}, P_{xx0}, P_{xxx})^T$, which is related to the desired set of occupation probabilities $\mathbf{P} = (P_{000}, P_{001}, P_{010}, P_{011}, P_{100}, P_{101}, P_{110}, P_{111})^T$ according to
\begin{eqnarray}
\mathbf{P}_{\mathrm{null}}
 & = &
  \left( \begin{array}{cccccccc}
    1 & 0 & 0 & 0 & 0 & 0 & 0 & 0 \\
    1 & 1 & 0 & 0 & 0 & 0 & 0 & 0 \\
    1 & 0 & 1 & 0 & 0 & 0 & 0 & 0 \\
    1 & 1 & 1 & 1 & 0 & 0 & 0 & 0 \\
    1 & 0 & 0 & 0 & 1 & 0 & 0 & 0 \\
    1 & 1 & 0 & 0 & 1 & 1 & 0 & 0 \\
    1 & 0 & 1 & 0 & 1 & 0 & 1 & 0 \\
    1 & 1 & 1 & 1 & 1 & 1 & 1 & 1
  \end{array}\right)
  \cdot\mathbf{P} \nonumber \\
 & = &
  \left( \begin{array}{cc} 1 & 0 \\ 1 & 1 \end{array}\right) \otimes
  \left( \begin{array}{cc} 1 & 0 \\ 1 & 1 \end{array}\right) \otimes
  \left( \begin{array}{cc} 1 & 0 \\ 1 & 1 \end{array}\right)
  \cdot\mathbf{P}.
  \label{xtalkfree}
\end{eqnarray}
By inverting this equation, we thus obtain the occupation probabilities in a way that is completely insensitive to measurement crosstalk.

Even with one qubit and no crosstalk, the measurement of the qubit state has finite fidelity, due to stray tunneling of the $\ket{0}$-state and relaxation of the $\ket{1}$-state during the measurement pulse.  Measured results can be corrected to account for this finite fidelity, and this correction is simpler than the crosstalk correction as the fidelity errors act independently on each qubit.  Hence the multi-qubit correction is just like single qubit case.  For each qubit, we determine the fidelity of $\ket{0}$-state measurement $f_0$ and the error probability of $\ket{1}$-state measurement $e_1 = 1-f_1$ from the so-called measurement ``s-curves''\cite{Lucero2008}.  Taking these into account, Equation \ref{xtalkfree} relating the measured null results to the actual occupation probabilities becomes
\begin{equation}
\mathbf{P}_{\mathrm{null}} =
  \left( \begin{array}{cc} f_0 & e_1 \\ 1 & 1 \end{array}\right)_A \otimes
  \left( \begin{array}{cc} f_0 & e_1 \\ 1 & 1 \end{array}\right)_B \otimes
  \left( \begin{array}{cc} f_0 & e_1 \\ 1 & 1 \end{array}\right)_C
  \cdot\mathbf{P}.
  \label{fidelity}
\end{equation}
Inverting this equation gives the occupation probabilities with single-qubit measurement fidelity taken into account.

The final subtlety in this measurement process is that because each element of $\mathbf{P}_{\mathrm{null}}$ is measured in separate repetitions of the experiment, each is subject to independent statistical noise.  As a result, when Equation \ref{fidelity} is inverted the elements of $\mathbf{P}$ may not be nonnegative and they may not sum to unity, as required for the set of probabilities.  We thus use maximum likelihood estimation (MLE)\cite{James2001} to enforce these constraints and find the probabilities $\mathbf{P}$ which would give the measured results $\mathbf{P}_{\mathrm{null}}$ with the highest probability.

\section{Quantum State Tomography}
\begin{figure*}
\includegraphics[trim=0in 4.2in 0in 0in, clip=true]{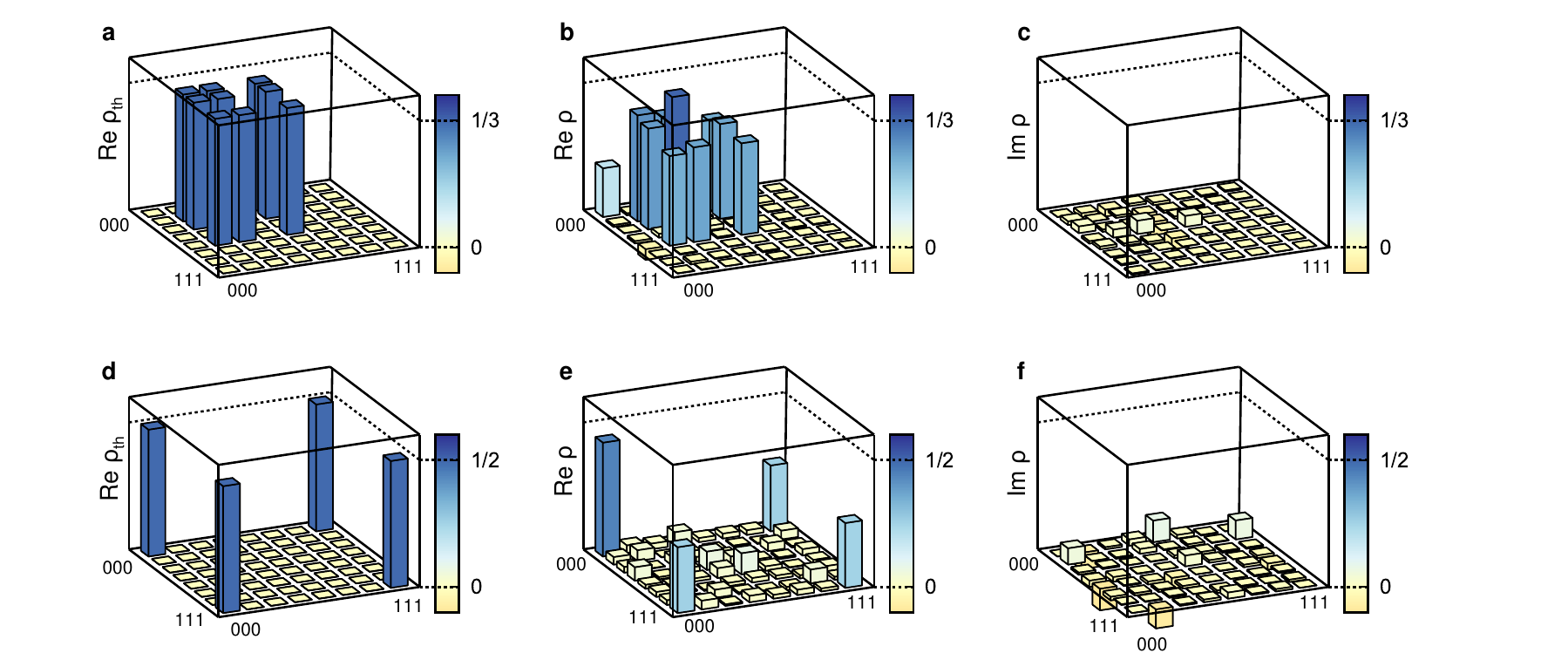}
\caption{\label{fig:matrices} Quantum state tomography of $\W$ and $\GHZ$.  {\bf a}, The real part of $\rho_{\mathrm{W}}^{\mathrm{th}} = \ket{\mathrm{W}}\bra{\mathrm{W}}$.  All nonzero elements are equal to $1/3$, and all imaginary parts (not shown) are identically zero.  The experimental $\rho_{\mathrm{W}}$ real part ({\bf b}) and imaginary part ({\bf c}) compare nicely with the theoretical prediction, giving fidelity $F_{\mathrm{W}} = 0.78$ with $|\Im\,\rho_{\mathrm{W}}| < 0.03$.  {\bf d}, The real part of $\rho_{\mathrm{GHZ}}^{\mathrm{th}} = \ket{\mathrm{GHZ}}\bra{\mathrm{GHZ}}$.  All nonzero elements are equal to $1/2$, and all imaginary parts (not shown) are identically zero.  The experimental $\rho_{\mathrm{GHZ}}$ real part ({\bf e}) and imaginary part ({\bf f}) again agree nicely with theory, giving fidelity $F_{\mathrm{GHZ}} = 0.62$ with $|\Im\,\rho_{\mathrm{GHZ}}| < 0.10$.}
\end{figure*}

To reconstruct the density matrix of the quantum state after the entangling protocol, we use a standard implementation of Quantum State Tomography\cite{James2001}.  After the state is prepared, we apply combinations of the identity operation or $\pi/2$-rotations about $X$ or $Y$ to each qubit before measuring the probabilities as outlined in the previous section.  MLE is used to extract a density matrix that gives the measured data with highest probability while also obeying the constraints on a physical density matrix: it must be Hermitian, positive semidefinite (no negative eigenvalues) and have unit trace.  If we forego the MLE procedure and instead use a straightforward least-square matrix inversion to perform state tomography, the resulting density matrix is typically Hermitian with unit trace, but may have one or two small negative eigenvalues (with magnitude on the order of 0.05 times the largest eigenvalue), indicating that these are likely due to independent statistical noise and finite fidelity of the individual measurement results, and not large systematic effects.  The theoretical and experimental density matrices are shown in Figure \ref{fig:matrices}.

\section{Characterizing entangled states}
\subsection{Entanglement Witnesses}
An entanglement witness is an operator that detects entanglement in quantum states, either pure or mixed.  We give here a few basic facts about witnesses, following closely the discussion in \cite{Acin2001}.  An operator $\mathcal{W}$ is an entanglement witness if $\Tr(\mathcal{W}\rho) \geq 0$ for {\it all} separable states $\rho$, but $\Tr(\mathcal{W}\rho) < 0$ for {\it some} entangled states $\rho$ in which case we say that the entanglement is ``detected'' by $\mathcal{W}$.  Note that such an entanglement witness defines a sufficient but not necessary condition for $\rho$ to be entangled, and in general there are many possible entanglement witnesses that each detect some subset of entangled states.  Some entanglement witnesses are constructed in such a way that they can be efficiently measured\cite{Toth2009}, allowing entanglement to be detected without performing full quantum state tomography which requires a number of measurements that grows exponentially with the number of qubits.  In the present experiment, however, we perform full state tomography on the three-qubit states, as the required number of measurements is still quite manageable, and then the measured density matrix $\rho$ can be checked against any desired entanglement witness.

The entanglement witnesses we use all have the form $\mathcal{W} = \epsilon\mathbf{1} - \ket{\psi}\bra{\psi}$ where $\ket{\psi}$ is some pure entangled state, $\mathbf{1}$ is the identity operator, and $\epsilon$ is a constant that depends on the state $\ket{\psi}$.  For such a witness, the detection condition reads $\Tr(\mathcal{W}\rho) = \epsilon - \bra{\psi}\rho\ket{\psi} < 0$ which then becomes $F_{\psi}\equiv\bra{\psi}\rho\ket{\psi} > \epsilon$ where $F_{\psi}$ is the fidelity of $\rho$ with respect to the state $\ket{\psi}$.  Hence, entanglement is detected if the fidelity is sufficiently high.  For the states of interest the relevant witnesses are\cite{Acin2001}
\begin{eqnarray}
  \label{eq:w}    \mathcal{W}_{\mathrm{W}} & = & \frac{2}{3}\mathbf{1} - \ket{\mathrm{W}}\bra{\mathrm{W}} \\
  \label{eq:ghz1} \mathcal{W}_{\mathrm{GHZ}_1} & = & \frac{1}{2}\mathbf{1} - \ket{\mathrm{GHZ}}\bra{\mathrm{GHZ}} \\
  \label{eq:ghz2} \mathcal{W}_{\mathrm{GHZ}_2} & = & \frac{3}{4}\mathbf{1} - \ket{\mathrm{GHZ}}\bra{\mathrm{GHZ}}.
\end{eqnarray}
We have changed the names of these witnesses from those given in the reference because our emphasis is slightly different.  For our purposes, $\mathcal{W}_{\mathrm{W}}$ is a witness that distinguishes $\ket{\mathrm{W}}$-like states from separable states and likewise $\mathcal{W}_{\mathrm{GHZ}_1}$ is a witness that distinguishes $\ket{\mathrm{GHZ}}$-like states from separable states.  Recalling from above that these witnesses are satisfied for fidelities greater than $\epsilon$, we see that the experimental fidelities $F_{\mathrm{W}} = \bra{\mathrm{W}}\rho_{\mathrm{W}}\ket{\mathrm{W}} = 0.78 > 2/3$ and $F_{\mathrm{GHZ}} \equiv \bra{\mathrm{GHZ}}\rho_{\mathrm{GHZ}}\ket{\mathrm{GHZ}} = 0.62 > 1/2$ both satisfy the respective entanglement witness.

The final witness operator $\mathcal{W}_{\mathrm{GHZ}_2}$ is a stronger condition that distinguishes the class of $\ket{\mathrm{GHZ}}$-like states from the class of $\ket{\mathrm{W}}$-like states.  The experimental fidelity $F_{\mathrm{GHZ}} < 3/4$ does not satisfy this witness, so that at least according to this criterion it is not possible to separate our $\rho_{\mathrm{GHZ}}$ from a convex combination of $\ket{\mathrm{W}}$-like states.  Recall however that entanglement witnesses always give sufficient but not necessary conditions for identifying entanglement.  A better measure of the tripartite entanglement of $\rho_{\mathrm{GHZ}}$ is provided by the Mermin-Bell inequality.

\subsection{Mermin-Bell inequality}
The $\ket{\mathrm{GHZ}}$ state was first discussed\cite{Greenberger1990} for its very strong quantum correlations which could rule out hidden variable models with a single measurement, rather than by taking many measurements and looking at correlations among them, as with Bell's Inequality.  However, this single-measurement violation would require an ideal pure state and perfect measurement fidelity, neither of which are experimentally feasible.  Mermin\cite{Mermin1990b} showed inequalities for the realistic mixed-state case that are obeyed by hidden variable models, but violated by $\ket{\mathrm{GHZ}}$ and its generalizations to higher numbers of qubits.  He also showed that the potential violation grows exponentially with the number of qubits.  For the three-qubit case considered here, classical hidden variable models must obey $G \equiv \expectation{XXX} - \expectation{XYY} - \expectation{YXY} - \expectation{YYX} \leq 2$, while the pure $\ket{\mathrm{GHZ}}$ state satisfies $G_{\ket{\mathrm{GHZ}}} = 4$.  Experimentally we find $G_{\rho_{\mathrm{GHZ}}} = 2.076$, which violates the inequality and hence rules out a hidden-variable model for the measurement correlations of the created state $\rho_{\mathrm{GHZ}}$.  This violation unambiguously separates the experimental state $\rho_{\mathrm{GHZ}}$ from the class of $\ket{\mathrm{W}}$-like states.

\bibliography{entanglement}